\documentclass[twocolumn,english,aps,prl,floatfix,amssymb,superscriptaddress]{revtex4-1}
\usepackage[T1]{fontenc}
\usepackage[utf8]{luainputenc}
\usepackage{color}
\usepackage{amsmath}
\usepackage{amssymb}
\usepackage{graphicx}
\usepackage{esint}
\usepackage{verbatim}

\makeatletter

\providecommand{\tabularnewline}{\\}

\@ifundefined{textcolor}{}
{%
 \definecolor{BLACK}{gray}{0}
 \definecolor{WHITE}{gray}{1}
 \definecolor{RED}{rgb}{1,0,0}
 \definecolor{GREEN}{rgb}{0,1,0}
 \definecolor{BLUE}{rgb}{0,0,1}
 \definecolor{CYAN}{cmyk}{1,0,0,0}
 \definecolor{MAGENTA}{cmyk}{0,1,0,0}
 \definecolor{YELLOW}{cmyk}{0,0,1,0}
}

\usepackage{babel}
\usepackage[bookmarks=false,linkcolor=blue,urlcolor=blue,colorlinks,citecolor=blue]{hyperref}
\@ifundefined{showcaptionsetup}{}{%
 \PassOptionsToPackage{caption=false}{subfig}}
\usepackage{subfig}
\makeatother

\usepackage{babel}
\begin{document}

\title{Emergent quasicrystals in strongly correlated systems}
\author{Eran Sagi}
\affiliation{Department of Condensed Matter Physics, Weizmann Institute of Science, Rehovot 76100, Israel}
\author{Zohar Nussinov}
\affiliation{Department of Physics, Washington University, St. Louis, MO 63130, U.S.A.}
\affiliation{Department of Condensed Matter Physics, Weizmann Institute of Science, Rehovot 76100, Israel}

\begin{abstract}
Commensurability is of paramount importance in numerous strongly interacting electronic systems.
In the Fractional Quantum Hall effect, a rich cascade of increasingly narrow plateaux appear at larger denominator filling fractions.
Rich commensurate structures also emerge, at certain filling fractions, in high temperature superconductors and other electronic systems. A natural question concerns the character of these and other electronic systems at {\it irrational} filling fractions. Here we demonstrate that quasicrystalline structures naturally emerge in these situations, and trigger behaviors not typically expected of periodic systems.  We first show that irrationally filled quantum Hall systems cross over into quasiperiodically ordered configuration in the thin-torus limit. Using known properties of quasicrystals, we argue that these states are unstable against the effects of disorder, in agreement with the existence of quantum Hall plateaux. We then study analogous physical situations in a system of cold Rydberg atoms placed on an optical lattice. Such an experimental setup is generally disorder free, and can therefore be used to detect the emergent quasicrystals we predict. We discuss similar situations in the Falicov-Kimball model, where known exact results can be used to establish quasicrystalline structures in one and two dimensions. We briefly speculate on possible relations between our theoretical findings and the existence of glassy dynamics and other features of strongly correlated electronic systems.
\end{abstract}

%\pacs{05.50.+q, 64.60.De, 75.10.Hk}
\maketitle

\maketitle

%{\it{Introduction.}}
The effects of commensurability appear in an extensive set of strongly correlated systems that, amongst many others, includes the high-temperature cuprate superconductors \cite{Tranquada1995,Yamada1998,Zaanen1989,Schulz1990,Kato1990,Emery1993} and the fractional quantum Hall systems \cite{Tsui1982,Laughlin1983}. These effects often arise from an intricate interplay between
the inherent length scales of the system and restrictions, e.g., those concerning magnetization or total particle number. Depending
on such externally imposed constraints, distinct phases may arise. This dependence
may have strong consequences for the nature of excitations and criticality in
such systems.

To motivate the quintessential physics investigated in this work, consider $N$ strongly interacting particles placed on the sites
of a periodic lattice of $M$ sites. In a continuum rendition of such a theory (i.e., one in which the particles do not need to occupy lattice sites),
when the interactions are long-ranged and repulsive, homogeneous Winger-crystal type \cite{Wigner1934} structures will be energetically preferred; the periodicity of such a Wigner lattice will be set by the density and the particle interactions.

In this study, the particles will be constrained
to reside on the discrete sites of another spatial structure -- the underlying periodic crystal. The latter lattice defining the theory may be incommensurate relative to
the basic periodicity of the ideal Wigner lattice. The mismatch
between the externally imposed lattice spacing and the energetically favored Wigner lattice length scale may spawn complex superlattice structures, which
may spontaneously break the original lattice translational symmetries (as well as
rotational and other point group symmetries).
\begin{figure}
\includegraphics[scale=0.6]{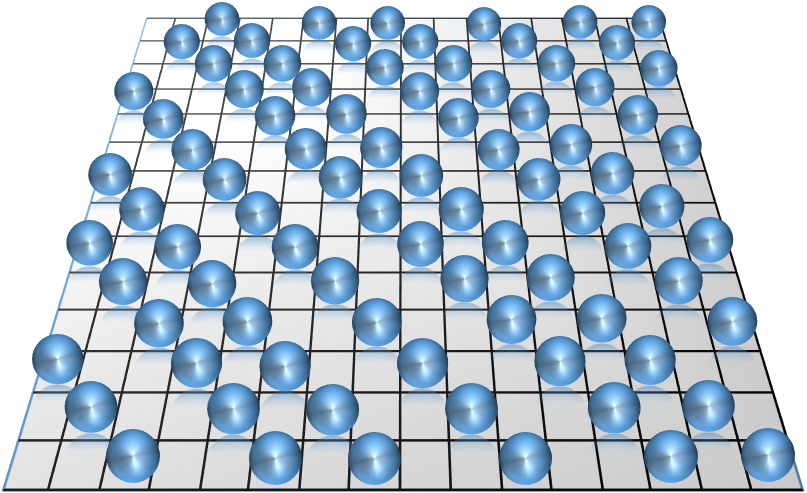}\protect\caption{\label{fig:structure}A finite patch of the electronic quasicrystal that emerges in the 2-dimensional Falicov-Kimball model. Notice that while the particles reside on a periodic lattice, the resulting structure must be aperiodic due to the irrational filling of the system. However, the strong interactions trigger a long range ordered quasiperiodic structure, as revealed by the point-like diffraction pattern.  Our results indicate that such quasicrystalline structures appear quite broadly in clean strongly correlated systems which exhibit the type of irrational frustration we describe. }
\end{figure}
In systems where the average number ($f=N/M$) of particles per site is irrational,
simple periodic order is prohibited. Nevertheless, long range interactions may still favor
the formation of structures with some form long-range correlations,
even in the absence of periodicity.

 As the above arguments hint, strongly interacting systems at
irrational filling fractions may exhibit rich structures (and ensuing physical characteristics).
Indeed, even deceptively simple-looking Ising and other
models \cite{Elliott1961, Bak1980, Fisher1980, Selke1988, Gendiar2005, Bak1982, Chayes1996, Giuliani2006, Chakrabarty2011, Giuliani2013, Giuliani2015} harbor a plethora of highly nontrivial ground states and dynamics \cite{Beijeren1990}
with devil staircase structures. Frustration effects between incommensurate length scales were additionally shown to induce incommesurately modulated crystals in various classical systems \cite{Frenkel1938,Janssen1995}.  It was further demonstrated experimentally that similar effects stabilize incommensurate composite crystals \cite{Lefort1996,Smaalen1996,Weber1996}. Related effects appear also in quantum systems, e.g., the frustration between the lattice and magnetic length scales in the Hofstadter problem \cite{Hofstadter1976} gives rise to a fractal spectrum \cite{Dean2013}.

As we will describe in this work, naturally occurring strongly correlated electronic systems and other quantum theories having only ubiquitous kinetic hopping, Coulomb, and spin exchange interactions may, quite broadly, harbor largely unexplored emergent quasicrystalline structures for irrational filling fractions (without spin-orbit terms explored in interesting recent studies \cite{Gopalakrishnan2013,Lifshitz2014,Sandbrink2014,Gopalakrishnan2014}). For instance, as we will demonstrate, even on periodic one- and two-dimensional ionic lattices, interactions may render
the underlying electronic structures to be quasicrystalline. The peculiar phenomenon of emergent quasicrystals (QCs) may trigger behaviors not typically expected of translationally invariant systems.

These new predicted {\it emergent} electronic (and other) quantum quasicrystalline structures
notably differ from quasicrystals discovered long ago in metallic alloys \cite{Shechtman1984,Levine1984,Levine1986} and intensely studied in the decades since. In the celebrated metallic alloy quasi-crystalline systems, the underlying ionic structure is, on its own, already quasi-periodic and may be further stabilized, in some cases, by electronic effects  \cite{Martin2015}. By contrast, in the systems
studied here an effective quasi-crystalline electronic (or other) structure emerges on periodic ionic or optical lattices.

%{\it{Quasicrystals and their symmetries.}}
In general, quasicrystals are aperiodic structures with well
defined Bragg peaks \cite{Shechtman1984,Levine1984,Levine1986}, i.e., the Fourier transform of the density takes the form
\begin{equation}
\rho(\mathbf{k})=\sum_{\mathbf{G}}\rho_{\mathbf{G}}\delta\left(\mathbf{k-G}\right), \label{eq:Diffraction pattern}
\end{equation}
where the reciprocal vectors are combinations of $D$ basis vectors, $\mathbf{G}=\sum_{i=1}^{D}m_{i}\mathbf{b}_{i},$ and the coefficients $m_{i}$ are integer valued. QCs differ from periodic crystals in that $D$ exceeds the spatial dimension $d$ \cite{Mermin1992,Lifshitz2011}.

A direct consequence of the above is that these structures have
$D$ global $U(1)$ symmetries, $\rho(\mathbf{G})\rightarrow\rho(\mathbf{G})e^{2\pi i\chi(\mathbf{G})}$, satisfying $\chi\left(\mathbf{G}=\sum_{i=1}^{D}m_{i}\mathbf{b}_{i}\right)=\sum_{i=1}^{D}m_{i}\chi(\mathbf{b}_{i}).$
Rigid translations are described by $d$ linear combinations of
$\chi(\mathbf{b}_{i})$. The remaining $(D-d)$ independent
phases describe additional global rearrangements
that generate distinct QCs with identical statistical
characteristics. These symmetry operations, having no analogs
in periodic crystals, are called phason symmetries \cite{Bak1985}.
Phason symmetries will play a key role in our investigation.

{\it{Quantum Hall systems at irrational filling factors.}} To concretely describe our results, we first study the quantum
Hall effect at an irrational filling and demonstrate the emergence of quasiperiodic structures. We predict that while these states are unstable against disorder, as evident from the observation of quantum Hall plateaux, one may, nonetheless, see signatures of the underlying quasiperiodic structure by looking at increasingly cleaner samples. Furthermore, as we will explain, similar phenomena appear in a system of cold Rydberg atoms, where disorder is absent and emergent quasicrystalline structures can be more crisply observed.

We focus on the so called ``thin-torus limit'', in which
the planar Quantum Hall (QH) system is mapped onto a one-dimensional (1D) classical problem \cite{Tao1983, Thouless1985,Tao1984,Bergholtz2005,Seidel2005,Bergholtz2008}.
The original two-dimensional (2D) fractional quantum Hall (FQH) Hamiltonian admits a natural 1D description within the guiding center representation, in which the single-electron
wavefunctions $\psi_{k}(\vec{r})$ are labeled by a 1D momentum momentum index $k$. In the Landau gauge, for example, the momentum $k$
in the $x$ direction controls the average position in
the $y$ direction.

Remarkably, when placing the FQH system on a torus for which the circumference associated with one of its directions (the ``$y$-direction'') far exceeds the system size along the other transverse (``$x$''-) direction, the
overlap between adjacent wavefunctions diminishes and the 1D model becomes classical.
Specifically, the
remnant non-vanishing terms lead to
\begin{equation}
\label{Hclass}
H=\sum_{j_{1}j_{2}}V(j_{1}-j_{2})n_{j_{1}}n_{j_{2}},%\label{eq:H in the thin torus}
\end{equation}
where the natural numbers $n_{j}$ denote the occupancies of respective states $\psi_{\frac{2\pi j}{L_x}}$. The projected interactions take the form
\begin{eqnarray}
V(j_{1}-j_{2})=\frac{1}{2}\int \int d^{2}r_{1}d^{2}r_{2} &&\Big[ \left| \psi_{\frac{2\pi j_1}{L_x}}(\vec{r}_{1})\right|^2 V_c(\vec{r}_{1}-\vec{r}_{2})\nonumber
\\ &&\times \left| \psi_{\frac{2\pi j_2}{L_x}}(\vec{r}_{2})\right|^2 \Big] ,\label{eq:V_k}
\end{eqnarray}
where $V_c(\vec{r}_{1}-\vec{r}_{2})$ is the coulomb
interaction.

Notwithstanding the formal simplicity of Eq. (\ref{Hclass}),
much nontrivial physics in captured by this classical Hamiltonian. Fortunately, a general solution
for this problem exists. Assuming a general repulsive interaction $V(j)$
which satisfies $V(j+1)+V(j-1)\geq2V(j)$ and vanishes as
$j\rightarrow\pm\infty$, a prescription for generating
the ground state configuration corresponding to any rational filling $f=\frac{p}{q}$
was presented in Ref. \cite{Hubbard1978}. This general recipe illustrates that the ground states are periodic, with a unit cell of size $q$. Examples of the ground state
configurations corresponding to various rational filling fractions $f$ are
provided in Table (\ref{tab:configurations}), the third column of which presents the pattern of consecutive $n_{j}$ values in a unit cell. We observe that in general, FQH states cross over into classical periodic states as we approach the thin-torus limit. Remarkably, some of the topological properties of the original FQH states, such as the fractional charges, are encoded in those periodic structures \cite{Tao1983, Thouless1985,Tao1984,Bergholtz2005,Seidel2005,Bergholtz2008}.
\begin{table}
\begin{tabular}{|c|c|c|c|}
\hline
$a$ & $f_{a}$ & Configuration & Sequence\tabularnewline
\hline
\hline
1 & $\frac{1}{2}=0.5$ & $10$ & $S$\tabularnewline
\hline
2 & $\frac{1}{3}=0.3333\cdots$ & $100$ & $L$\tabularnewline
\hline
3 & $\frac{2}{5}=0.4$ & $10100$ & $SL$\tabularnewline
\hline
4 & $\frac{3}{8}=0.375$ & $10010100$ & $LSL$\tabularnewline
\hline
5 & $\frac{5}{13}\approx0.38461\cdots$ & $1010010010100$ & $SLLSL$\tabularnewline
\hline
6 & $\frac{8}{21}\approx0.38095\cdots$ & $100101001010010010100$ & $LSLSLLSL$\tabularnewline
\hline
\end{tabular}\protect\caption{\label{tab:configurations}The ground state configurations corresponding
to 1D systems whose filling factors are given by the sequence $f_{a}=F_{a}/F_{a+2}$,
where $F_{a}$ is the $a$'th Fibonacci number. The first and second
columns present the index $a$ and the corresponding value of $f_{a}$.
The third column presents the unit cell of the ground state configurations
in the occupation basis, and the fourth column presents these in the
compact notation $S=10$ and $L=100$. }
\end{table}

Thus far, we largely reviewed the properties of the ground state in the thin torus limit. We now explicitly turn to our new results associated with irrational filling factors $f$. As an illustrative example specifically associated with
Table (\ref{tab:configurations}), we set $f$ equal to an archetypal
irrational number, $f=1-\tau=\frac{3-\sqrt{5}}{2}\approx0.38197$, where $\tau$ is
the reciprocal of the golden-ratio, $\tau=\frac{\sqrt{5}-1}{2}$.

 A sequence of rational numbers that converges to $(1-\tau)$
is provided by $f_{a}=\frac{F_{a}}{F_{a+2}}$, where $F_{a}$ is the
$a$'th Fibonacci number ($F_{a}=F_{a-1}+F_{a-2}$ with $F_{1}=F_{2}=1$).
Using the general prescription of Ref. \cite{Hubbard1978},
we may then generate the periodic ground state configurations for any such
fraction.
The unit cells corresponding to $a=1,2,\cdots,6$ are presented
in Table (\ref{tab:configurations}). Pursuing Table (\ref{tab:configurations}), one observes that two adjacent occupied sites are always separated by either
one or two empty sites. We verified that this persists for
higher values of $a$ as well. The ground state configurations can
therefore be compactly encoded by combinations of the strings $S=10$
and $L=100$ as presented in the last column of Table (\ref{tab:configurations}).

Interestingly, we find that the $(a+1)-$th unit cell may be iteratively generated
from the $a-$th cell via the inflation rules $S\rightarrow L$,
$L\rightarrow SL$. Remarkably, these are the very same inflation rules defining
the Fibonacci QC \cite{Senechal1996,Lifshitz2002}. To verify
that the ground state configuration tends to that of the Fibonacci QC
in the irrational $f_{a}$ limit, we numerically compute the Fourier transform
of the function $n(j)$ corresponding to $a=15$ (for which the
unit cell is of length $1597$).
\begin{figure}
\includegraphics[scale=0.55]{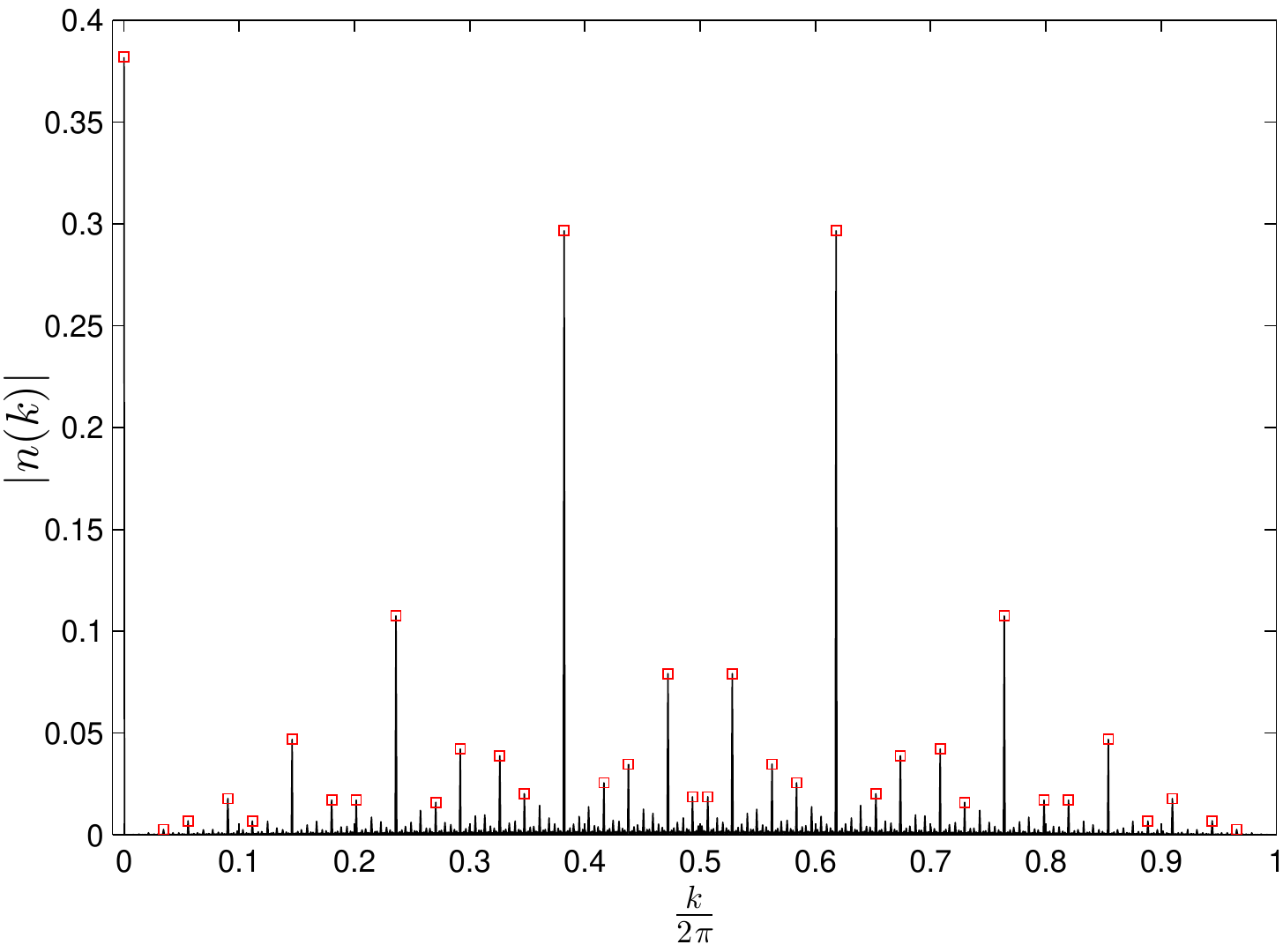}\protect\caption{\label{fig:Fourier transform}The Fourier transform of the occupation
in the ground state configuration corresponding to $a=15$, whose
unit cell is of size 1597. For comparison, the red squares represent the Fourier components
of the Fibonacci quasicrystal. }
\end{figure}
In Fig. (\ref{fig:Fourier transform}), these Fourier weights are contrasted with the Fourier components of the Fibonacci QC
(whose Bragg peaks are located at $k=2\pi\left[\left(\tau l\right)\text{mod}1\right]$, where $l$ is an integer, and Fourier weights can easily be calculated analytically \cite{sagi2014}).
Aside from small amplitude fluctuations, which asymptotically tend to zero as $a$
increases, the two diffraction patterns coincide to a very
good approximation.

The above numerical evidence indicates that in the $a\rightarrow\infty$ limit, the ground state configuration
coincides with the Fibonacci QC. The realization of a QC structure has immediate physical consequences. This is so
as QCs exhibit the earlier noted continuous $U(1)$ phason symmetries. These symmetries become transparent when writing the Fourier components of the density in the form
\begin{equation}
n(k)=\sum_{l=-\infty}^{\infty}\delta(k-2\pi l\tau)n_{l}e^{2\pi il\chi} \label{eq:Fourier transform of density},
\end{equation}
where the phason symmetry is manifest as a $U(1)$ invariance under changes of the phase $\chi$.

{\it Low energy excitations for general interactions.} We now describe the low lying excitations about QC ground states for disparate potentials $V$. Such excitations result from effective long-range spatial
variations of the phase $\chi$. However, as $\chi$ is a globally
defined quantity, any such description
poses a fundamental difficulty. Heuristically, however, we may partition
the system into large patches whose linear spatial size is still much smaller
than the scale of change in $\chi$. As $\chi$ is essentially
a constant on the scale of a single patch, it can be defined locally
by calculating the Fourier-transform of the density in that region.
This intuitive idea can be implemented formally with the aid of the
Local Fourier Transform (LFT) \cite{sagi2014},
\begin{equation}
n(k,j)=\frac{1}{A}\sum_{m}w_{\sigma}(j-m)n(m)e^{-ikm},\label{eq:LFT}
\end{equation}
where $w_{\sigma}(m)$ is a weight function that is equal to unity in a region
of linear size $\sigma$ centered around the origin, and vanishes
otherwise. The parameter $A$ is defined as $A=\sum_j w_\sigma (j)$. Pictorially, $n(k,j)$ indeed describes the Fourier transform
of $n(j)$ inside a patch of size $\sigma$, centered around the point
$j$. In terms of the LFT, we can write the low energy excitations as
\begin{equation}
n(j,k)=\sum_{l=-\infty}^{\infty}\delta(k-2\pi l\tau)n_{l}e^{2\pi il\chi(j)},\label{eq:LFT of quasicrystal}
\end{equation}
 where $\chi(j)$ is a slowly varying function $\sigma \partial_{x} \chi \ll1$. Writing the Hamiltonian in terms of the Local Fourier components, and invoking Eq. (\ref{eq:LFT of quasicrystal}), we obtain

\begin{figure}
\includegraphics[scale=0.5]{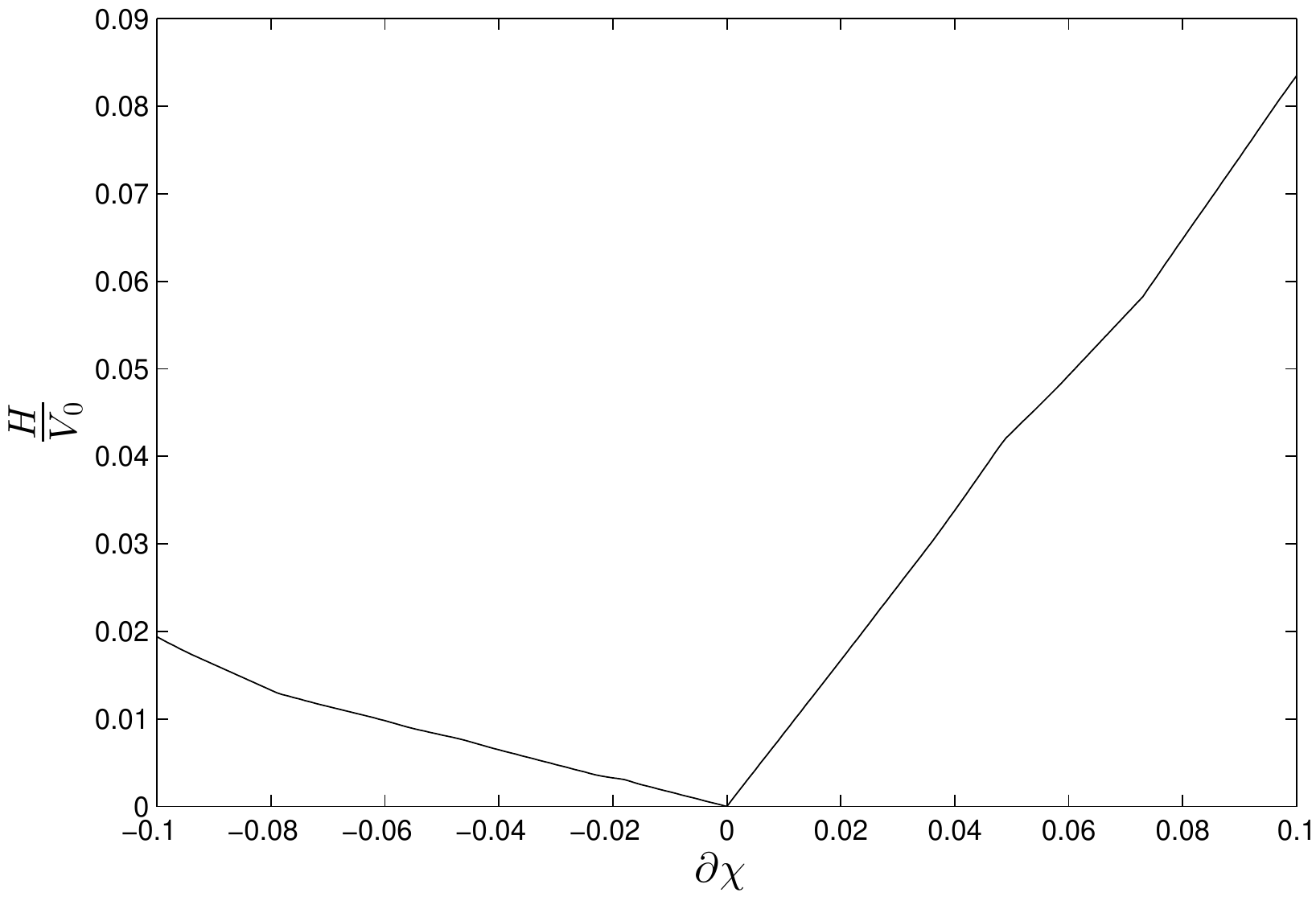}\protect\caption{\label{fig:energy_alpha_3}Phason excitation energies as a function of $\partial_{x}\chi$, for $V(m)=V_{0}m^{-3}$}
\end{figure}
\begin{equation}
H=\sum_{m}V(m)\sum_{l,j}\left|n_{l}\right|^{2}e^{2\pi iml\left(\partial_{j}\chi(j)+\tau\right)}. \label{eq:plugging slow excitations}
\end{equation}
Here, we approximated $\chi(j+m)-\chi(j)=m\partial_{j}\chi(j)$.
The sum over $l$ can be performed analytically, as shown in the supplemental material \cite{Supp}.

For a given interaction $V$, the energy associated with the low-energy configurations
can be evaluated directly from the resulting expression.
To illustrate this, we plot in Fig. (\ref{fig:energy_alpha_3}),
the energy corresponding to $V(m)=V_{0}m^{-3}$. For this interaction (and other potentials of the form $V(m)=V_0 m^{-\alpha}$ \cite{Supp}), we obtain a non-analytic energy
dependence of the form
\begin{equation}
H=\sum_{j}\left|\partial_{j}\chi(j)\right|\left[a+b\Theta(\partial_{j}\chi(j))\right],\label{eq:H}
\end{equation}
for small $\partial_{j}\chi$. Thus, in the limit of strong interactions the system is gapless,
in agreement with the Goldstone theorem.

As our primary model is one dimensional,
the quantum fluctuations caused by inflating the torus inhibit spontaneous breaking of the continuous phason symmetry. Instead, the system
may exhibit algebraic correlations of the form $\left\langle e^{2\pi i\left[\chi(j)-\chi(0)\right]}\right\rangle \propto |j|^{-\eta}$. Such a behavior was observed numerically in Ref.  \cite{sagi2014} in an analogous 2D classical system.

More generally, as our results indicate that the FQH states at irrational filling are
gapless, these states are unstable to disorder. Such an instability is consistent with the
existence of quantum Hall plateaux. In fact, we can easily use the
above framework to establish the hierarchy of quantum Hall
states, in which the gap (and therefore, the region of stability)
of quantum Hall states of filling $\nu=p/q$ monotonically decreases with
$q$. This is shown in details in the supplemental material \cite{Supp}.

Despite the above instability with respect to disorder, it is clear that as the quality of samples improves, additional Quantum Hall plateaux emerge, serving as approximants to the underlying quasicrystalline patterns.

{\it Quasicrystals in an ultra-cold atomic system.}  Following Refs. \cite{Sela2011, Weimer2010}, we now study a system of cold Rydberg atoms placed on an optical lattice (of arbitrary spatial dimensionality). These disorder free systems are natural candidates for observing the emergent quasicrystalline structures that we find. We assume that each site contains exactly one particle. Using an external laser, a transition from the ground state to an excited Rydberg state is enabled, making each lattice site a two-level system. We note that a realization of such a setup was reported in Ref. \cite{Schauss2012}.

We label the states by a pseudo-spin $S_z = \uparrow / \downarrow$ (where $\downarrow$  represents the ground state and $\uparrow$ represents the excited state). In terms of the spin-1/2 degrees of freedom, the Hamiltonian can be written in the form  \cite{Sela2011}:
\begin{eqnarray}
H & = & \sum_{\mathbf{R},\mathbf{R'}}J(\left|\mathbf{R-R'}\right|)\left(S_{\mathbf{R}}^{z}+\frac{1}{2}\right)\left(S_{\mathbf{R'}}^{z}+\frac{1}{2}\right)\label{eq:Hamiltonian cold atoms} \\
 &  & -\frac{J_{\perp}}{2}\sum_{\left\langle \mathbf{R,R'}\right\rangle }\left(S_{\mathbf{R}}^{+}S_{\mathbf{R'}}^{-}+h.c.\right)+\sum_{\mathbf{R}}\left(\Omega S_{\mathbf{R}}^{x}-\Delta S_{\mathbf{R}}^{z}\right).\nonumber
\end{eqnarray}
Here, $\Omega$ is the Rabi frequency, $\Delta$ is the detuning, $J(\left|\mathbf{R}\right|)=\frac{J_0}{\left|\mathbf{R}\right|^\alpha}$ represents the repulsive interactions between excited atoms (which can be, e.g., of the dipole-dipole type, for which $\alpha=3$), and the parameter $J_\perp$ quantifies the hopping of excitations. In realistic experimental setups the largest scale is $J_0$, prompting us to start by neglecting all other terms. Once we do that, it is clear that the ground state is polarized, with $S_{\mathbf{R}}^{z}=-1/2$ for all $\mathbf{R}$. If we introduce a non-zero positive $\Delta$, it becomes energetically preferable to have a finite density of up-spins.

In fact, for any rational number $f=p/q$, we can find a finite range of $\Delta$ for which the density of up-spins is $f$ in the ground state. Notice, however, that the size of this range diminishes as $q$ increases. Taking the long-range interactions into account, and starting from the 1D situation for simplicity, the considerations used to study quantum Hall systems can be replicated {\it mutatis mutandis}, showing that the up-spins form periodic structures with a unit cell of size $q$ (see Table \ref{tab:configurations}).

In particular, specializing to the sequence $f_a$ of densities, the resulting unit cell is of size $F_{a+2}$, and converges again to the Fibonacci quasicrystal as the index $a$ increases. For any value of $a$, there are $F_{a+2}$ distinct ground states differing by translations. For convenience, the different ground states are labeled by the parameter $\chi \equiv \left(d \frac{F_{a+1}}{F_{a+2}}\right)\text{mod}1$, where $d$ is an integer describing the translation with respect to some reference ground state. The parameter $\chi$ can take the values $0,\frac{1}{F_{a+2}},\frac{2}{F_{a+2}},\cdots,\frac{F_{a+2}-1}{F_{a+2}}$, and uniformly covers the segment $[0,1)$ in the limit $a\rightarrow\infty$. This notation is useful as in the limit $a\rightarrow\infty$, the parameter $\chi$ represents the phason symmetry of the Fibonacci quasicrystals. We represent a ground state configuration as $u_a^\chi u_a^\chi u_a^\chi u_a^\chi u_a^\chi u_a^\chi \cdots$, where $u_a^\chi$ is the unit cell (for example, for $f=1/2$, we get $u_1^0=10$ and $u_1^{1/2}=01$).

For any finite $a$, the low energy excitations include configurations that differ from a ground state by a set of domain walls. These can generally be represented as $u_a^{\chi_1} u_a^{\chi_2} u_a^{\chi_3} u_a^{\chi_4} u_a^{\chi_5} \cdots$, where the label $\chi$ varies in space. Following Eq. (\ref{eq:H}), we model the energy of such a configuration as $H_C=\sum_{n}\phi(\chi_n-\chi_{n+1})$, where $\phi$ is a periodic function which coincides with Eq. (\ref{eq:H}) for small changes. Next, we can consider the effects of quantum fluctuations induced, e.g., by a non-zero $J_\perp$.

We verified numerically that the smallest change in $\chi$ within a unit cell is reduced in real space to an exchange of spin between adjacent sites. Consequently, the term multiplying $J_\perp$ connects states of different $\chi$. Assigning a state written in the basis $\left\{ \left|\chi=0\right\rangle ,\left|\chi=\frac{1}{F_{a+2}}\right\rangle ,\left|\chi=\frac{2}{F_{a+2}}\right\rangle \cdots\left|\chi=\frac{F_{a+2}-1}{F_{a+2}}\right\rangle \right\}$ in each unit cell, we get a term of the form
$H_Q=\frac{J_{\perp}}{2}\sum_{n}\sigma_{n}+h.c.,$ where the operators $\sigma_n$ are defined such that $\sigma_n \left|\chi_n\right\rangle=\left|\left(\chi_n+\frac{1}{F_{a+2}}\right)\text{mod}1\right\rangle$.
 We refer to the resulting 1D model (given by $H_C+H_Q$) as a modified quantum $F_{a+2}$-state clock model, in which the energy associated with spatial variations of the spin is determined by the function $\phi$.

 This model is in the universality class of the 2D classical clock model, which evolves into that of the 2D classical $XY$ model in the limit $a\rightarrow \infty$. We conclude that arbitrarily weak quantum fluctuations demote the quasicrystalline ground state into a quasi-long-range ordered phase with algebraic correlations of the form $\langle\exp\left[2\pi i\left(\chi_n-\chi_{n+m}\right)\right]\rangle\propto m^{-\eta(J_\perp,J_0)}$. However, following the numerical results presented in Ref. \cite{sagi2014}, we may speculate that $\eta$ is typically parametrically small. Therefore, while true quasicrystalline long-range-order cannot exist in 1D, Bragg peaks can still be observed in small systems.

 In higher spatial dimensions ($D>1$), quasicrystalline long range order is expected to survive the introduction of quantum fluctuations. Indeed, in what follows we turn to study the Falicov-Kimball model, in which exact results may be used to demonstrate the validity of our arguments beyond 1D.

{\it The Falicov-Kimball model.} We now use similar considerations to study the irrationally filled Falicov-Kimball model.
The Hamiltonian \cite{Falicov1969}
\begin{equation}
H=-t\sum_{\left\langle \mathbf{R,R'}\right\rangle }\left(f_{\mathbf{R}}^{\dagger}f_{\mathbf{R'}}+h.c.\right)+U\sum_{\mathbf{R}}c_{\mathbf{R}}^{\dagger}c_{\mathbf{R}}f_{\mathbf{R}}^{\dagger}f_{\mathbf{R}},\label{eq:FK}
\end{equation}
portrays interactions between spineless itinerant and localized electrons.
Here, $f_{\mathbf{R}}$ ($c_{\mathbf{R}}$) is the annihilation operator of itinerant (localized)
electrons at site $\mathbf{R}$.
Alternatively, the $c$- (or $f$-) fermions may portray positively (negatively) charged
ions (electrons) with an attractive (i.e., $U<0$) interaction. We assume that the total electron and ion numbers are equal and
fixed, $\frac{1}{N}\sum_{\mathbf{R}}f_{\mathbf{R}}^{\dagger}f_{\mathbf{R}}=\frac{1}{N}\sum_{\mathbf{R}}c_{\mathbf{R}}^{\dagger}c_{\mathbf{R}}=f$.

In the
limit of large negative $U$, the second term drives
localized electron-ion bound states.
The second term, on the other hand, favors electron delocalization, and therefore acts qualitatively as repulsive interactions between bound states. In 1D, for any rational $f=p/q$, and sufficiently large $|U|$ \footnote{Following \cite{Lemberger1992}, in 1D,  the threshold minimal value ($|U_{\min}|$) of $|U|$ required to stabilize period $q$ ground state is bounded by $|U_{\min}|  < c \times 4^{q}$
with $c$ a constant.}, the ions form a periodic lattice
with a unit cell of size $q$ \cite{Lemberger1992}. In two-dimensions (2D) \cite{Haller,Kennedy}, a periodic arrangement of diagonal stripes emerges
for rational $f\in[\frac{1}{3},\frac{2}{5}]$, wherein the stripe locations assume configurations identical to those in the corresponding
ground states of the 1D system.  A prescription for constructing
the 1D ground state configurations was provided in \cite{Lemberger1992}.

These earlier rigorous results pave the way for our exact study of the irrationally
filled model in 1D (and 2D). As before, we set $\frac{1}{3} < f=(1-\tau) < \frac{2}{5}$ in
the large $|U|$ regime \footnote{As, formally, the minimal $|U_{\min}|$ required to enable a proof of the period $q$ stripe order may diverge as $q \to \infty$, we may consider the electron density $f$ to be a highly incommensurate fraction with a large denominator $q$ that corresponds to one of the $a \gg 1$ elements in the set of rational sequence approximants $f_{a} = \frac{F_{a}}{F_{a+2}}$, with $F_{a}$ the $a$-th Fibonacci number.}. Approximating $f$
with high denominator elements of the rational number sequence $\{f_{a}\}$, as we have in the systems described above, and invoking the results
of Ref. \cite{Lemberger1992} (and \cite{Kennedy,Haller}), we discover structures identical to those
found in the Fibonacci QC. Putting all of the pieces together, {\it quasicrystalline type order} may emerge for large denominator approximants to irrational particle densities $f$, in
{\it two-dimensional} electronic systems. A sketch is provided in Fig. (\ref{fig:structure}).

As we argued above, in this case, quantum fluctuations are not expected to destabilize the quasicrystalline nature of the model. Thus, the electronic structure
may, similar to a periodic crystal, reveal sharp Bragg peaks for momenta parallel to the direction of the stripes and concomitantly exhibit more intricate quasicrystalline features for momenta transverse to the stripe direction. If quantum and/or thermal fluctuations or disorder partially suppress the quasicrystalline features then the resulting momentum space patterns may be qualitatively similar to those anticipated for ``electronic liquid crystals''
(in particular for those of the nematic type)  \cite{kiv,seamus}.

{\it Conclusions.}
We demonstrated that
QC type ground states and associated gapless excitations appear in a broad set of one- and two-dimensional strongly
interacting systems. Clearly, disorder, fluctuations, and other effects may stabilize
more standard commensurate orders (or destroy these altogether). One may, nevertheless, expect to find imprints of the underlying quasiperiodic structures even when these are destroyed, e.g., in the form of stable approximants with a finite (but large) unit cell as experimentally appears elsewhere \cite{kelton}. Alternatively, these effects may result in non-homogeneous systems containing puddles of approximate quasiperiodic structures.

As an immediate consequence of our results, we expect these systems to be associated with slow dynamics, due to the exceptionally long relaxation of the phason degrees of freedom. In particular, one may postulate that the very slow dynamics observed in some correlated electronic systems via NMR and NQR may be rationalized by phason-type excitations.

 In general, quasicrystals may exhibit certain features similar to those of structural glasses such as stretched exponential type dynamics \cite{stretch}.  Specifically, the structural relaxation in an equilibrated quasicrystal is composed of an initial rapid (so-called $\beta$ type) relaxation, which is followed by a slower ($\alpha$ type) relaxation with a stretched exponential behavior as in glasses. This behavior is generally associated with the phason degrees of freedom. In particular, when supercooled from high temperatures, a system that is a quasicrystal in equilibrium might become quenched into a glass just as more common supercooled liquids do. Furthermore, in metallic liquids, compositions that lead to glasses and quasicrystals often lie in close proximity to each other \cite{kelton1,bookQCs}. In fact, certain theories consider glasses to be aperiodic crystals \cite{Lubchenko2007}.

Taken together, all of the above suggest that systems exhibiting quasicrystalline ground states may be unstable to (i) commensurate lock-in effects (possibly to high order approximants in clean systems) or, as underscored above, (ii) an inherent susceptibility towards glassy dynamics and aperiodic structures. Indeed, in certain strongly correlated electronic and disorder free magnetic systems, stretched exponential decay and other features of glassy (or possible other extremely slow) dynamics appear \cite{curro,Tuson,pnas2015,triangleus,Ovadia2015}.

 The fate of the emergent quasicrystals that we found theoretically and imprints thereof including, notably, possible relations between our prediction of electronic and atomic QCs to experimental findings remain to be tested by numerics.

\begin{acknowledgements}{\it Acknowledgements.}
We thank Seamus Davis, Eli Eisenberg, Iliya Esin, Ron Lifshitz, Yuval Oreg, Gerardo Ortiz, Alexander Seidel, Eran Sela, and Dan Shahar for insightful discussions. We acknowledge financial support from the NSF DMR-1411229, the Feinberg foundation visiting faculty program at the Weizmann Institute, and the Adams Fellowship Program of the Israel Academy of Sciences and Humanities.

\end{acknowledgements}

\bibliographystyle{apsrev4-1}
%\bibliography{ref}
%merlin.mbs apsrev4-1.bst 2010-07-25 4.21a (PWD, AO, DPC) hacked
%Control: key (0)
%Control: author (72) initials jnrlst
%Control: editor formatted (1) identically to author
%Control: production of article title (-1) disabled
%Control: page (0) single
%Control: year (1) truncated
%Control: production of eprint (0) enabled
%

\onecolumngrid
\newpage

\section*{\large{Supplemental Material}}

\subsection*{Energy of phason excitations for general interactions }

In this part, we show that the sum over $l$ in Eq. (7) of the main
text can be performed analytically, leaving us with a closed form,
from which we can get the energy for any interaction $V(m)$.

We start from the sum presented in the text:
\begin{equation}
H=\sum_{m}V(m)\sum_{l,j}\left|n_{l}\right|^{2}e^{2\pi iml\left(\partial\chi(j)+\tau\right)}.\label{eq:H}
\end{equation}
The coefficients $n_{l}$ can be evaluated analytically (see Appendix
C of {[}Phys. Rev. E 90, 012105{]}), and take the form
\begin{equation}
n_{l}=\delta_{l,0}-\frac{\sin\left[\pi\tau l\right]}{\pi l}.\label{eq:nl}
\end{equation}
Plugging this into $H$, and performing the sum over $l$, we finally
get the result

\[
H=\sum_{m}V(m)\left[\tau^{2}+\frac{2\text{Li}_{2}\left(e^{-2\pi im(\tau+\partial\chi)}\right)-\text{Li}_{2}\left(e^{-2\pi im(\frac{m-1}{m}\tau+\partial\chi)}\right)-\text{Li}_{2}\left(e^{-2\pi im(\frac{m+1}{m}\tau+\partial\chi)}\right)}{4\pi^{2}}\right.
\]
\begin{equation}
\left.+\frac{2\text{Li}_{2}\left(e^{2\pi im(\tau+\partial\chi)}\right)-\text{Li}_{2}\left(e^{2\pi im(\frac{m-1}{m}\tau+\partial\chi)}\right)-\text{Li}_{2}\left(e^{2\pi im(\frac{m+1}{m}\tau+\partial\chi)}\right)}{4\pi^{2}}\right].\label{eq:Li}
\end{equation}

\subsection*{Gap of rational approximants}

As noted in the main text, we can use the framework we developed to
establish the hierarchy of quantum Hall states, in which the gap of
quantum Hall states of filling $\nu=p/q$ monotonically decreases
with $q$.

To study this, we reexamine the sequence $f_{a}$ of rational filling
factors. Generating the corresponding periodic structures, and defining
$\tau_{a}\equiv1-f_{a}=F_{a+1}/F_{a+2}$, it is clear that we have
a finite number of independent reciprocal lattice vectors (i.e., reciprocal
vectors which do not differ by terms of the form $2\pi n$), which
can be written as
\begin{equation}
G_{m}=2\pi\tau_{a}m,\label{eq:G_m}
\end{equation}
where $0\le m\le F_{a+2}-1$ is an integer.

The Fourier transform can therefore be written in the form
\begin{equation}
n(k)=\sum_{l=0}^{F_{a+2}-1}\delta(k-2\pi\tau_{a})n_{l}e^{2i\pi\chi},
\label{eq:FTofapproximant}
\end{equation}
where the phase $\chi$ takes the values $\chi(j)=\left(\tau_{a}d\right)\text{mod}1$,
and $d$ is an integer. Shifting $d$ by any integer corresponds to
a rigid translation of the entire lattice.

For sufficiently large values of $a$, we may examine slow spatial
changes of $\chi$. In terms of the LFT, this is expressed as
\begin{equation}
n(j,k)=\sum_{l=0}^{F_{a+2}-1}\delta(k-2\pi l\tau_{a})n_{l}e^{2\pi il\chi(j)}.\label{eq:FT of approximant-1}
\end{equation}
Following the strategy presented in the main text, the energy of such
a configuration takes the form
\begin{equation}
H=\sum_{m}V(m)\sum_{l,j}\left|n_{l}\right|^{2}e^{2\pi il\left(\chi(j+m)-\chi(j)+\tau_{a}\right)}.\label{eq:plugging slow excitations-1}
\end{equation}

The excitation whose energy is minimal is expected to be a domain
wall of the form $\chi(j)=\chi_{0}+\Theta(j)\frac{1}{F_{a+2}}$. Once
this form is plugged in Eq. (\ref{eq:plugging slow excitations-1}),
and the values of $\left|n_{l}\right|$ are determined numerically,
we obtain the energy gap as a function of $a$.
\begin{figure}
\includegraphics[scale=0.7]{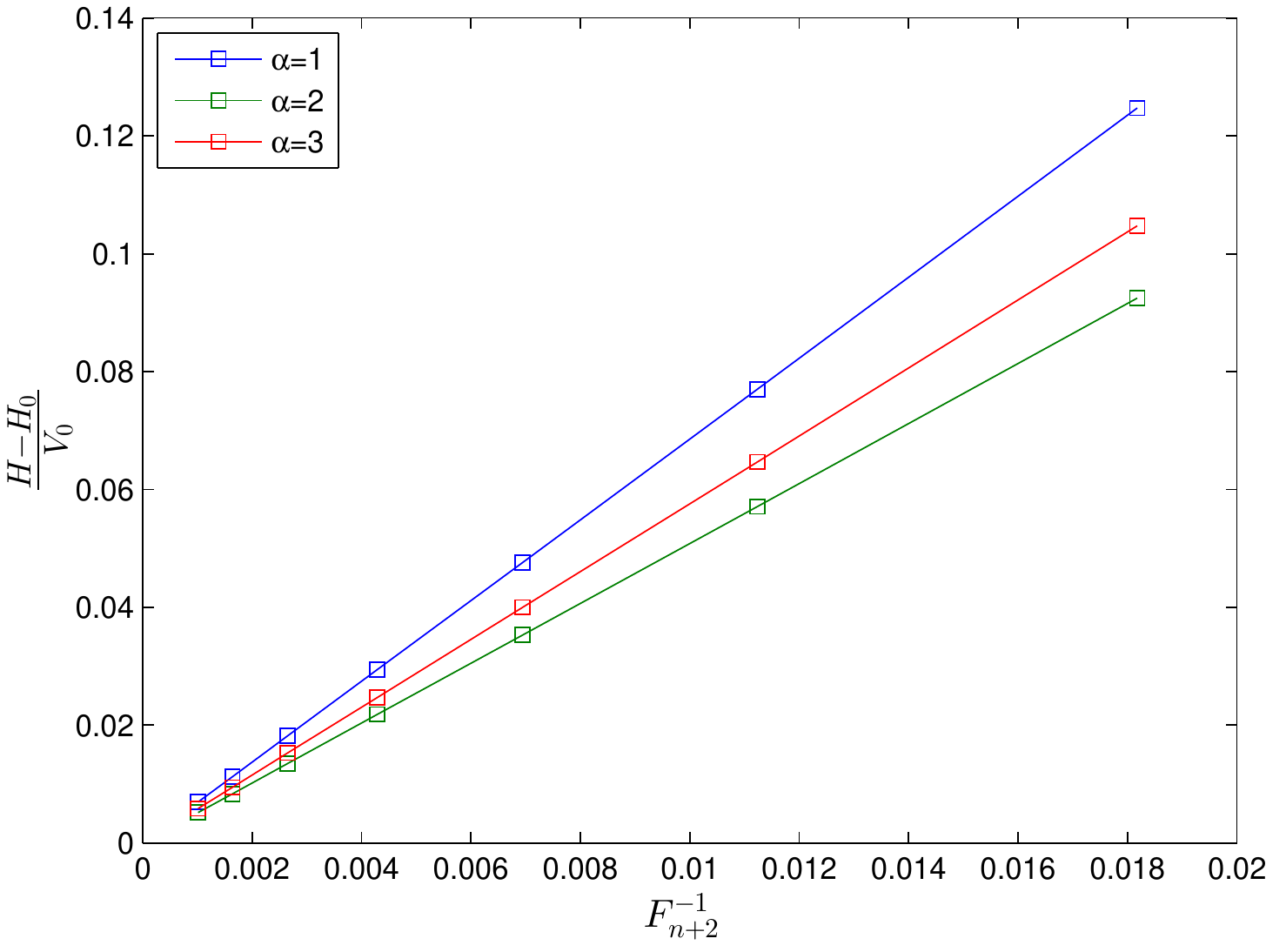}

\protect\caption{\label{fig:gap}The gap to excitations as a function of $F_{a+2}^{-1}$
(at fillings $\frac{F_{a}}{F_{a+2}}$) for the interactions $V(m)=V_{0}m^{-\alpha}$
($\alpha=1,2,3$).}
\end{figure}

The results of this analysis are presented in Fig. (\ref{fig:gap}),
where the energy gap associated with filling $f_{a}=\frac{F_{a}}{F_{a+2}}$
is plotted as a function of $F_{a+2}^{-1}$. The linear dependence
indicates that the gap scales like $F_{a+2}^{-1}$, and in particular,
tends to zero in the irrational limit, as shown explicitly in the
main text.

\subsection*{Excitation energy for various interactions }

\begin{figure}
\subfloat[\label{fig:a}]{\includegraphics[scale=0.82]{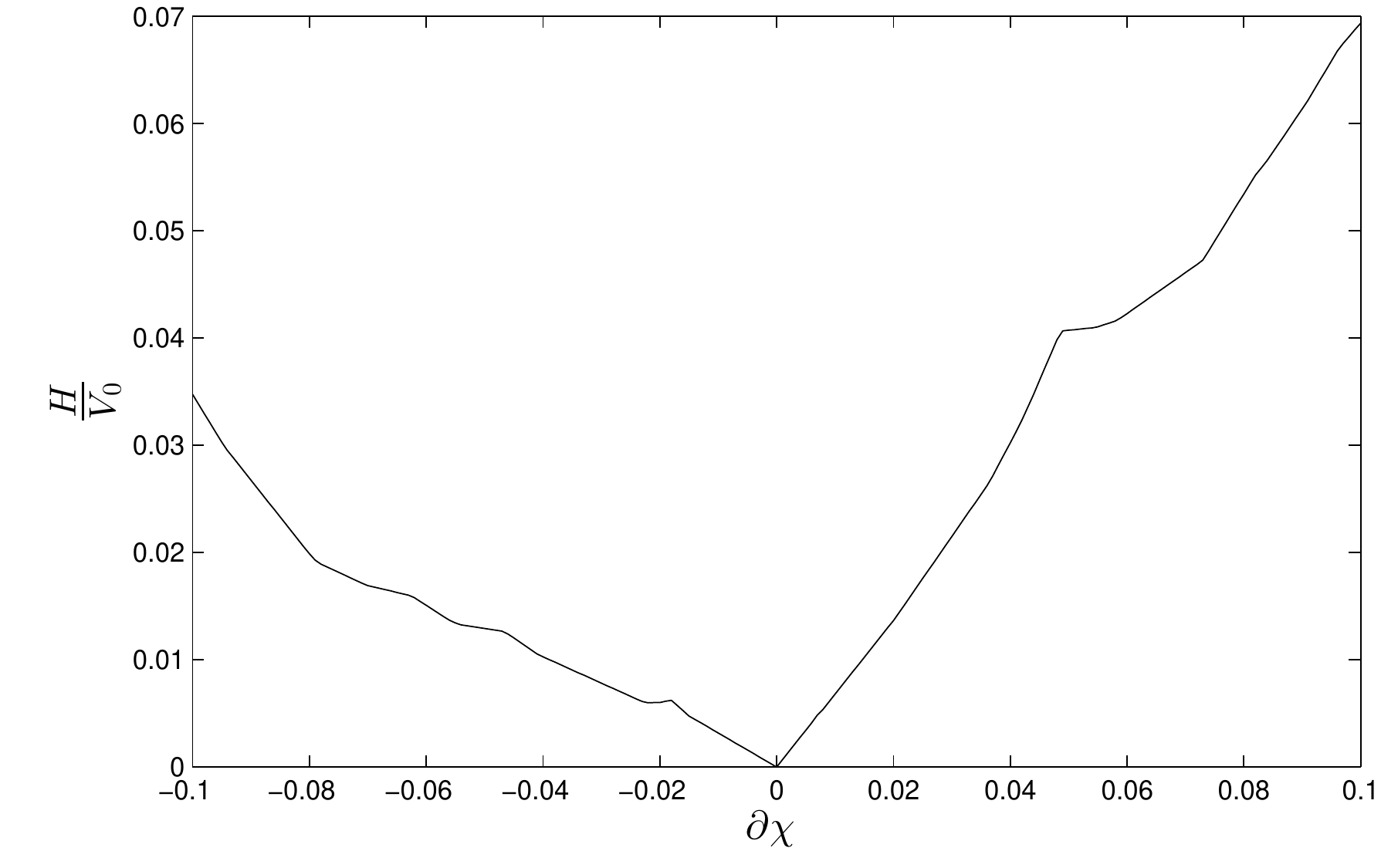}

}

\subfloat[\label{fig:b}]{\includegraphics[scale=0.8]{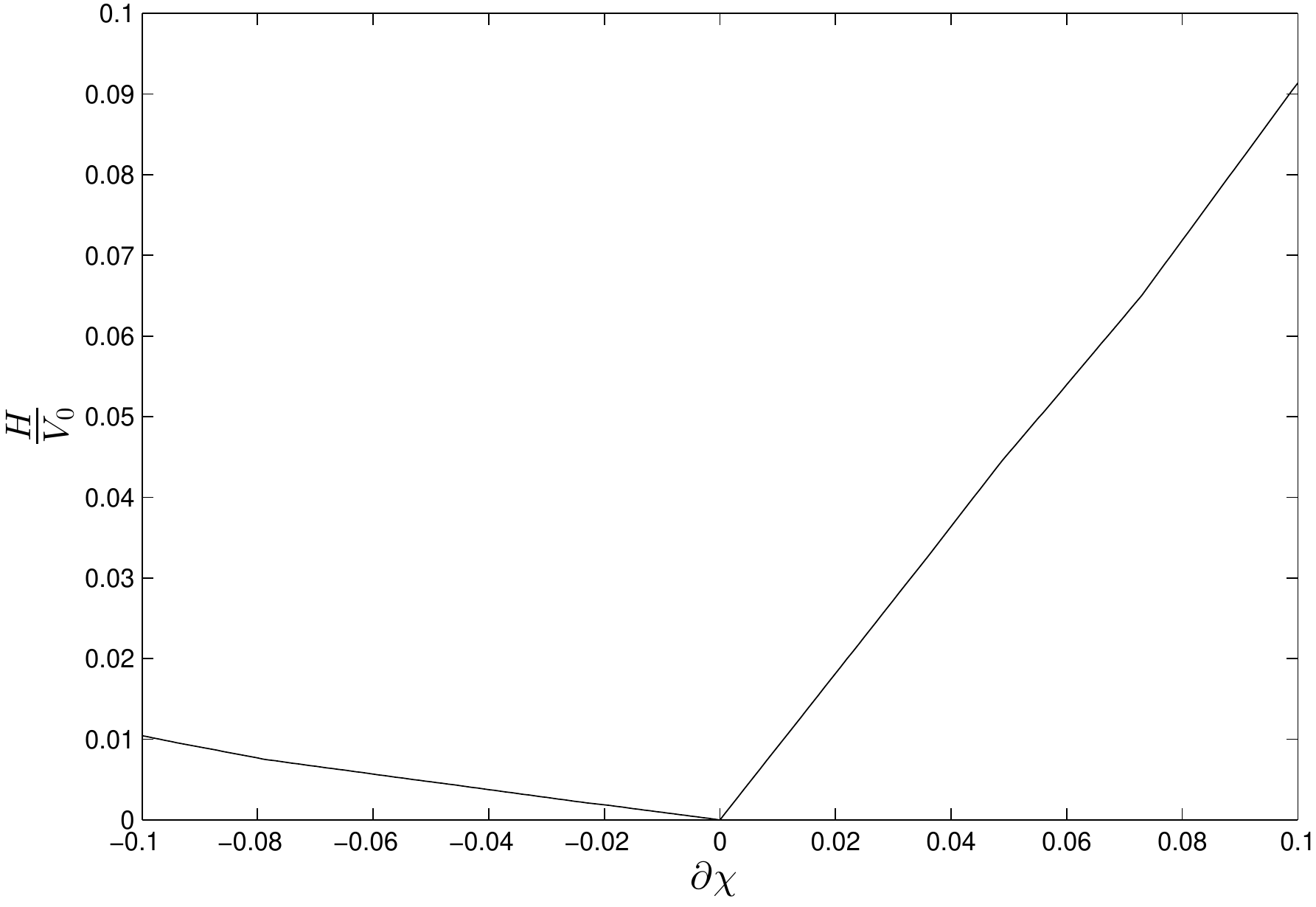}

}

\protect\caption{\label{fig:energy} The energy as a function of $\partial\chi$ for
$V(m)=V_{0}m^{-\alpha}$ with (a) $\alpha=2$ (b) $\alpha=4$.}
\end{figure}

We argued in the main text that the low-energy Hamiltonian
\begin{equation}
H=\sum_{j}\left|\partial\chi(j)\right|\left[a+b\Theta(\partial\chi(j))\right]\label{eq:energy}
\end{equation}
applies universally to all potentials we have checked. To illustrate
this result, we present in Fig. (\ref{fig:energy}) the energy as
a function of $\partial\chi$ for $V(m)=V_{0}m^{-\alpha}$ for $\alpha=2,4$
(in addition to the case $\alpha=3$ presented in the main text).

\end{document}